\documentclass[
amsmath,amssymb,
aps, physrev,
pra,
]{revtex4-2}
\usepackage{lipsum, babel}
\usepackage[utf8]{inputenc}
\usepackage{graphicx}
\usepackage{mathrsfs}
\usepackage{placeins}
\usepackage{rsfso}
\usepackage{hyperref}
\usepackage{amsmath}
\usepackage{subcaption}
\usepackage{dcolumn}
\usepackage{bm}
\usepackage{xcolor}
\usepackage{float}   
\usepackage[normalem]{ulem} 
\begin{document}

\title{Inverted Leakage Radiation Microscopy for measurement of nonreciprocity induced by a plasmonic metasurface with false chirality} 
\author{Ahmed Lafeef Ettapuram Naduvilepurayil}
\author{Sahil Sahoo}
\affiliation{Department of Electrical and Electronics Engineering, Ariel University, Ariel 40700, Israel}
\author{Yuri Gorodetski}
\email[]{yurig@ariel.ac.il}
\affiliation{Department of Electrical and Electronics Engineering, Ariel University, Ariel 40700, Israel}
\affiliation{Department of Mechatronics and Mechanical Engineering, Ariel University, Ariel 40700, Israel}
\begin{abstract}
A system has a nonreciprocity when the light propagating through it in opposite directions encounters different optical adventures.
This effect lies in the origin of various applications, including signal processing, noise reduction, unidirectional propagation, and sensing.
Here, we experimentally demonstrate the detection of nonreciprocity in plasmonic metasurfaces and a method to determine its direction. We use the Inverted Leakage Radiation Microscopy setup to show that the non-reciprocity is linked to a false chirality of the medium. 
The setup behaves as an equivalent of the classical Sagnac loop providing an interference pattern at the image plane. The non-reciprocity is then detected by examining the topological order of the resulting image. Our system represents a compact and versatile method to detect false chirality in plasmonics and suggests a variety of potential areas of implementation in nanophotonics, chemistry and applied physics. 
\end{abstract}
\maketitle 
\section{Introduction}
A system has a nonreciprocity when the light that propagates through it in opposite directions encounters different optical adventures\cite{caloz2018electromagnetic}.
Theoretically, this can be explained based on time-reversal (TR) symmetry breaking, parity of physical quantities, and various reciprocity theorems\cite{onsager1931reciprocal,onsager1931reciprocal2,caloz2018electromagnetic, caloz2018nonreciprocity, asadchy2020tutorial}. 
Nonreciprocity arises only when the TR symmetry is broken, which can occur in a system that is designed such that its response remains unchanged for time-odd quantities under TR.\cite{caloz2018electromagnetic, caloz2018nonreciprocity, asadchy2020tutorial}
Various phenomena like the Faraday\cite{faraday1933faraday4}, Sagnac\cite{sagnac1914effet}, Fresnel drag\cite{huidobro2019fresnel}, and Kerr polarization\cite{kaplan1981enhancement, kaplan1982directionally, kaplan1983light} effects exhibit nonreciprocity and are critical for applications requiring unidirectional light propagation.

Systems with false chirality\cite{barron1986true} inherently exhibit nonreciprocity as they break TR symmetry.
Previously\cite{ettapuram2025direct}, we showed that a plasmonic Pancharatnam–Berry (PB) metasurface, consisting of spatially rotated rectangular apertures featuring false-chirality, is a nonreciprocal system.
Surface plasmons were excited via a Kretschmann configuration.
The interaction between plasmons and the metasurface introduced a PB phase that broke the symmetry between counter-propagating modes, which led to a measurable nonreciprocal phase shift.
This effect was experimentally detected and studied in detail by implementing a custom Sagnac interferometer.
We confirmed the results by comparing them with nonchiral reciprocal structures (flat surface and hole array structure).
Moreover, the plasmonic dispersion modes measured in the $k$-space shed light on the link between the nonreciprocity and false chirality.

While the integration of a plasmonic interaction within a standard Sagnac loop was a straight-forward solution it has demonstrated some significant drawbacks. Specifically it has shown a very high sensitivity to misalignments and required a very fine parameter tuning and a cumbersome system. In addition, due to the setup limitations, it could only detect the nonreciprocity between plasmonic waves propagating forward or back along a predefined direction (i.e., $\pm k_x$ directions).

In this paper, we replace the bulky Sagnac setup by more compact and convenient plasmonic system -- inverted leakage radiation microscopy (ILRM).
This setup uses a single objective lens for both the excitation of the plasmons on the sample and for the collecting and re-collimating of the reflected beam. The emerging signal bears the signature of the plasmonic resonance encoded in the amplitude and a phase modulation of each plane-wave component. The image formed in the conjugate plane to the sample shows the characteristic interference pattern of all the reflected plane-waves from which the nonreciprocity can be assessed. Moreover, separate plane-wave components can be analyzed using a $k$-space imaging using an additional Fourier lens.    
Accordingly, the ILRM setup can be considered as a Sagnac interferometer for counter-propagating planewaves incident from opposite radial positions at the plasmonic medium.
The nonreciprocal structures are characterized by a second-order vortex as a base interferogram when illuminated and analyzed via opposite circular polarizers.
A change from this pattern reveals the presence of nonreciprocity in the structure.

An additional important feature of this setup is that, due to the radial arrangement of the illuminating plane-wave components, the reciprocity/nonreciprocity of the sample can be detected in any direction of the entire 2D plane.
Moreover, by a slight adjustment of the objective one can access the pre-focal position which allows the setup to visualize the diffraction orders of the structure (this is shown in the Supplementary Material).
The appearance of the nonreciprocity is then shown to be directly linked to the asymmetric diffraction pattern resulting from a geometric PB phase. 
Nonreciprocity detection using plasmonic metasurfaces may provide a pathway toward next-generation nanophotonic devices, with potential applications in amplitude and phase modulations\cite{mahmoud2015all,yang2023self}, and spatiotemporal light manipulation \cite{barati2022optical,shaltout2019spatiotemporal,guo2019nonreciprocal,sahoo2026spatio}. 

\section{Experimental Results and Discussions}

\begin{figure}[htbp]
    \centering
    \includegraphics[width=1\textwidth]{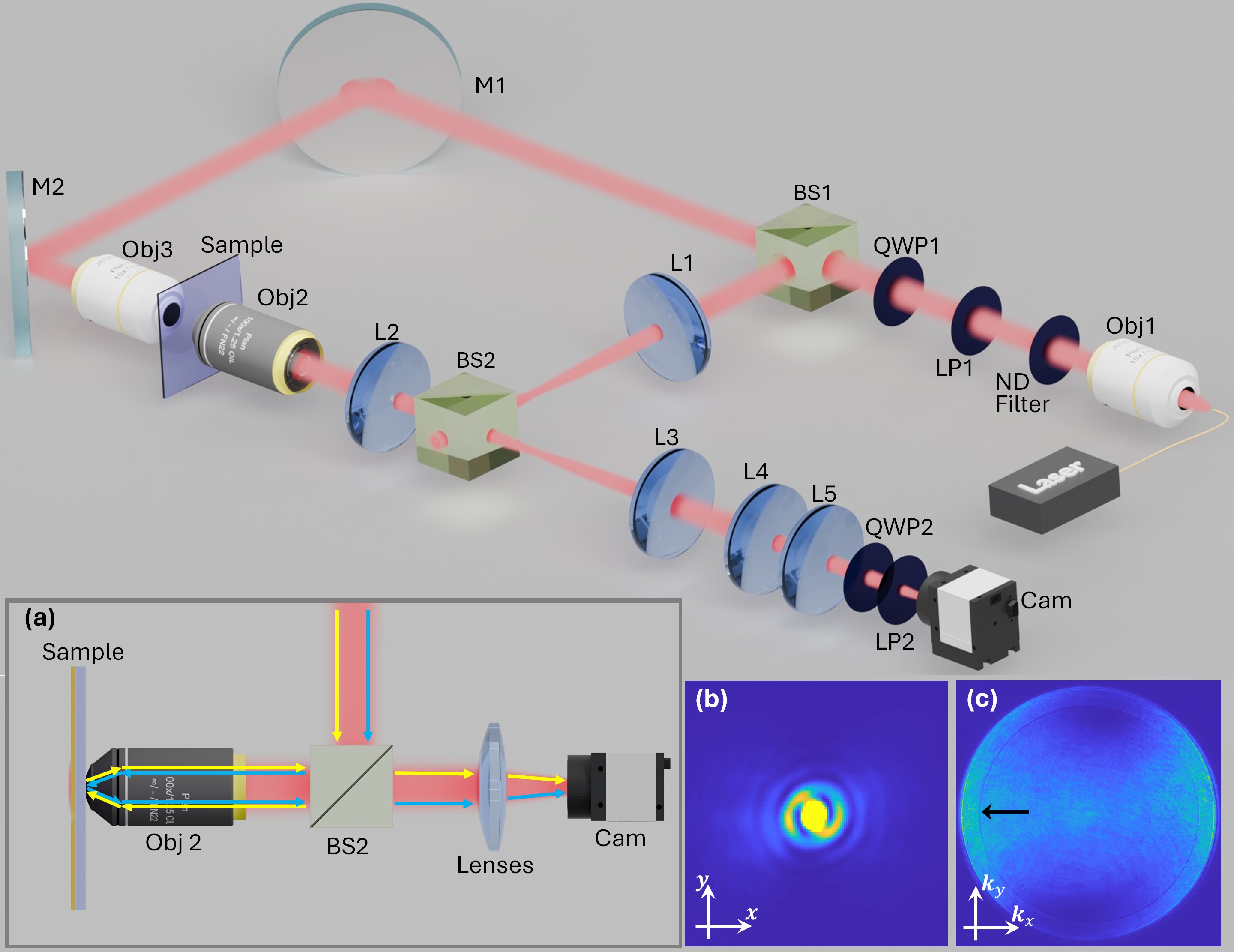}
    \caption{The combined Inverted - Direct Leakage Radiation Microscopy Setup. (a) A simplified scheme of the ILRM setup; more details in the text. (b) The real space image formed by the interference of the beams propagating in opposite directions on the sample. (c) The $k$-space image showing the dispersion mode due to the resonant excitation of Surface Plasmons along a flat surface of the sample.
    \label{fig:ILRM Setup}}
\end{figure}

\begin{figure}[htbp]
    \centering
    \includegraphics[width=0.4\textwidth]{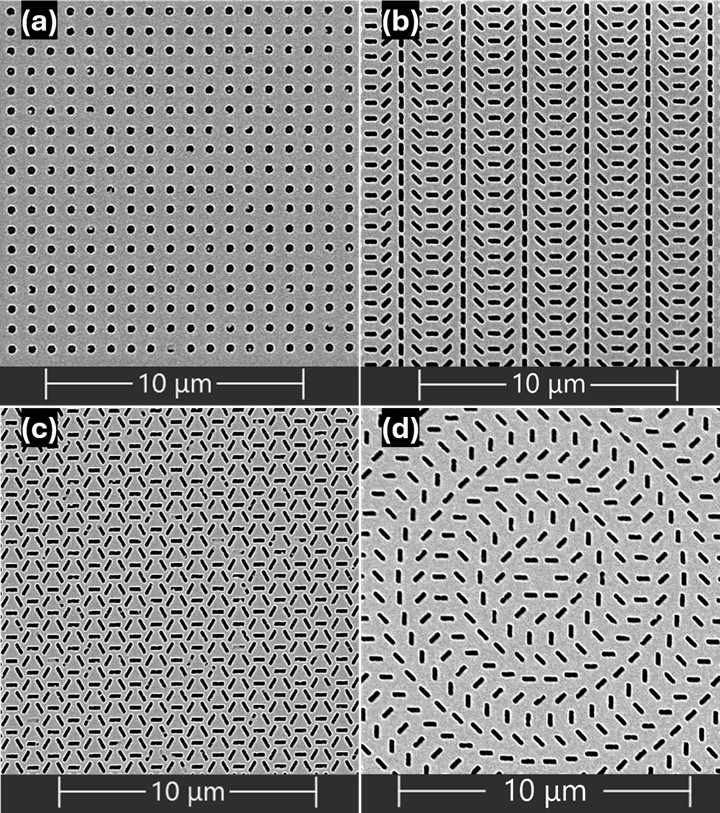}
    \caption{SEM image of samples studied. Reciprocal structure (nonchiral): (a) Hole Array (HA) gratings in a square lattice. Nonreciprocal structures (false-chiral): (b) Metasurface of shorter period ($\Lambda_{s}$) with spatially rotated rectangular apertures along the $x$ axis in a square lattice. (c) Metasurface of spatially rotated rectangular apertures along the three directions in a hexagonal lattice. Reciprocal structure (true-chiral): (d) Metasurface of rectangular apertures of same orientation at same radial position in a circular lattice. The rotation at each subsequent radial position is $\pi/4$ with N = 4.
    \label{fig: Samples}}
\end{figure}

The scheme of the experimental setup is depicted in Fig. \ref{fig:ILRM Setup}. A diode laser ($\lambda_0 = 785$ nm) was collimated using an objective lens (Obj1).
The laser beam then passed through a linear polarizer (LP1) and a quarter wave plate (QWP1) to ensure the required circular polarization (CP) state.
The beam-splitter (BS1) divided the beam into a ``direct leakage'' and an ``inverted leakage'' paths.
In the direct leakage radiation microscopy configuration, the beam was focused onto the gold sample from the air side using an objective lens (Obj3, 10x, NA = 0.25).
In the ILRM configuration, the beam was focused onto the sample from the glass side using a high numerical aperture objective lens (Obj2, 100x, NA = 1.25) attached to the sample with index matching oil.
The same high NA objective lens (Obj2) collected the leakage radiation from the plasmonic sample in both configurations.
In order to utilize one configuration, the unnecessary path was blocked.
The emerging beam was then transmitted through a tube lens (L2) and a couple of relay lenses (L3,L4) and the final image was recorded with a CMOS camera (Basler a2A2448-75umBAS).
We used an additional Fourier lens (L5) to observe the $k$-space.
A quarter wave plate (QWP2) and a linear polarizer (LP2) were used to filter out the initial circular polarization. In the following study we have tested a series of plasmonic structures with different geometries (see Fig. \ref{fig: Samples}).

The Sagnac loop obtained in the ILRM is presented schematically in Fig. \ref{fig:ILRM Setup}a.
Here we consider the incident collimated beam as a bundle of parallel rays with a circular cross section. Rays at opposite radial positions traverse opposite loops between the incidence at the sample and the image plane. The example of such a ray pair is marked by a blue and a yellow arrows in the Figure. By a proper setting, the optical paths of these rays are perfectly balanced, so that the interference appearing in the image plane at the camera reveals the system's nonreciprocity. Clearly, this analysis can be applied to any couple of  radially opposite rays.
In our system, the use of an objective lens with index-matching oil extends the incident NA to 1.25, allowing for a cone of rays that includes the plasmonic momentum, $k_{SP} = k_0\sqrt{\frac{\epsilon_m}{\epsilon_m+\epsilon_d}}$, where $k_0 = 2\pi/\lambda_0$, $\epsilon_m$ and $\epsilon_d$ are the free space incident wavenumber, the dielectric constant of the metal (Au) and the dielectric constant of the substrate (SiO), respectively.
This radially expanding surface wave then leaks back into a glass substrate and propagates through the optical system to the camera.
In order to filter out the direct reflection from the gold surface we use a circular polarizer/analyzer set with opposite polarization states. For instance, when a right-handed circular polarization (RCP) impinges on the sample, we observe the output beam through a left-handed circular (LCP) state analyzer and vise-versa. It has been shown that a spin-orbit interaction during plasmonic wave excitation is responsible for the excitation of a helical plasmonic wavefront of the form $E_{SP} \propto [\cos \varphi, \sin\varphi]^T e^{\pm i\varphi}$ (with $\varphi$ being an azimuthal angle) \cite{gorodetski2009obs}. Here, the sign of the phase depends on the light handedness, and the local orientation of the electric field vector is represented by the radial polarization. The beam passing through the circular analyzer is then expected to be decorated with a double spiral geometrical phase $\phi = \sigma2\varphi$ where $\sigma = \pm1$ is a spin number defining the handedness of the circular polarization state\cite{gorodetski2009obs}. In other words, the base image obtained after the light interaction with SP waves on a perfect flat gold surface represents a second-order optical vortex with a handedness corresponding to the incident light spin. Hereafter, we refer to this interference pattern as a signature of the reciprocal system. In the case where the interaction with the surface structure breaks the TR symmetry, the ILRM setup should benchmark the nonreciprocity. 
Importantly, due to the topological nature of the arising geometric phase, the system becomes very sensitive even to a minute nonreciprocity\cite{guo2019nonreciprocal}.

\begin{figure}[htbp]
    \centering
    \includegraphics[width=0.9\textwidth]{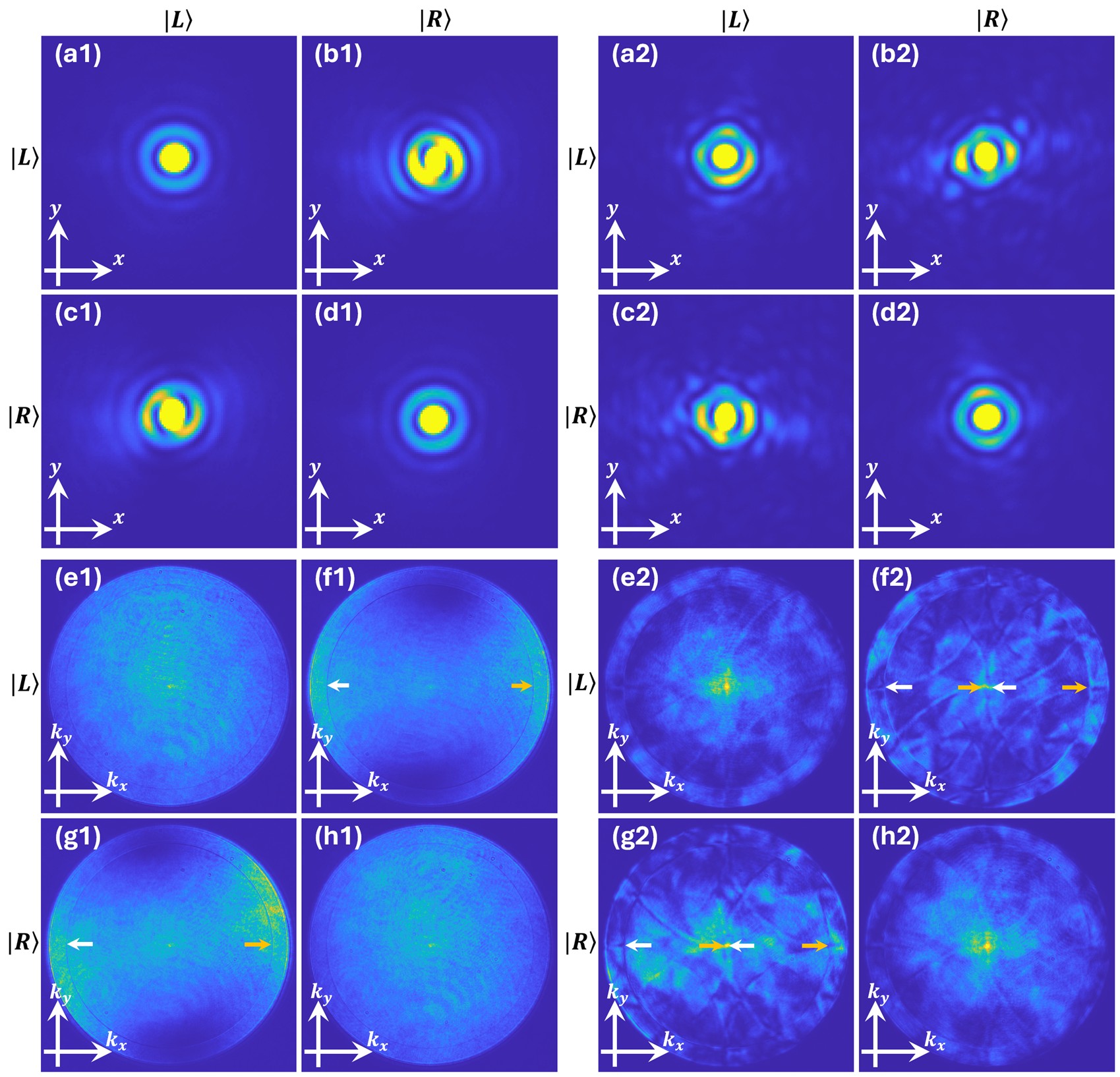}
    \caption{Results for Reciprocal Structures: (a1) - (h1) flat surface, (a2) - (h2) HA grating.
    The incident and measured circular polarizations are denoted on the left side and upper side of the figure, respectively, as $|L\rangle$ for LCP and $|R\rangle$ for RCP.
    Upper half represents real space results while the lower half shows $k$-space results with dispersion modes.
    \label{fig: Reciprocal_Flat_HA}}
\end{figure}

The experimental results measured for a flat surface are presented in  Fig. \ref{fig: Reciprocal_Flat_HA}a1-h1. This sample has been prepared by sputtering a $60\:nm$ gold layer onto a $160\:\mu m$ thick glass cover slip.
By a careful comparison of the intensity structures one can recognize the double vortex structure for crossed polarized configurations $ \left | R\right > \rightarrow \left | L\right >$ and $ \left | L\right >\rightarrow \left | R\right >$ (here $ \left | R/L\right >$ stands for the RCP/LCP states). When observing the light reflected from the sample at the same polarization state, a clear zeroth-order Bessel-type spot is observed indicating the absence of a geometric phase. As expected from the flat surface the SPs propagating in opposite radial direction do not break the time-reversal symmetry, so this system is reciprocal. We also investigated the $k$-space image obtained with L5 as shown in Fig.\ref{fig:ILRM Setup}. In the images taken with a crossed polarized configuration, a dark circle with a radius of $k_{SP}$ is clearly observed. This line confirms the excitation of SP waves in the radial direction. Interestingly, two darker spots are observed on the radially opposite sides on the line, manifesting the two singularities of the emerging vortex beam.

Next, we study an example of a reciprocal plasmonic structure - hole array (HA) in a square lattice with a period of $\Lambda = 765\:nm$ (see Fig. \ref{fig: Samples}a). This structure was fabricated using a focused ion beam (FIB, Helios 5). The results are presented in Fig. \ref{fig: Reciprocal_Flat_HA}a2-h2. Again, we note a specific signature of a double vortex with the crossed polarization configuration, though this time, due to the polarization cross-coupling the image is less pronounced. Inspection of the $k$-space reveals additional plasmonic modes that appear as replicas of the fundamental plasmonic resonance due to the Bloch-Floquet theorem\cite{floquet1883equations,bloch1929quantenmechanik}. We mark the secondary resonances with arrows to guide the eye. We note that the structure period chosen such as $k_g = 2\pi/\Lambda = k_{SP} $ results in an additional coupling of normally incident light to the SP. This leads to the brightest point in the center of the $k$-space and well pronounced secondary resonant lines. Nevertheless, we observe a perfect symmetry in any radial direction, a fact indicating the reciprocity of the system.

\begin{figure}[htbp]
    \centering
    \includegraphics[width=0.9\textwidth]{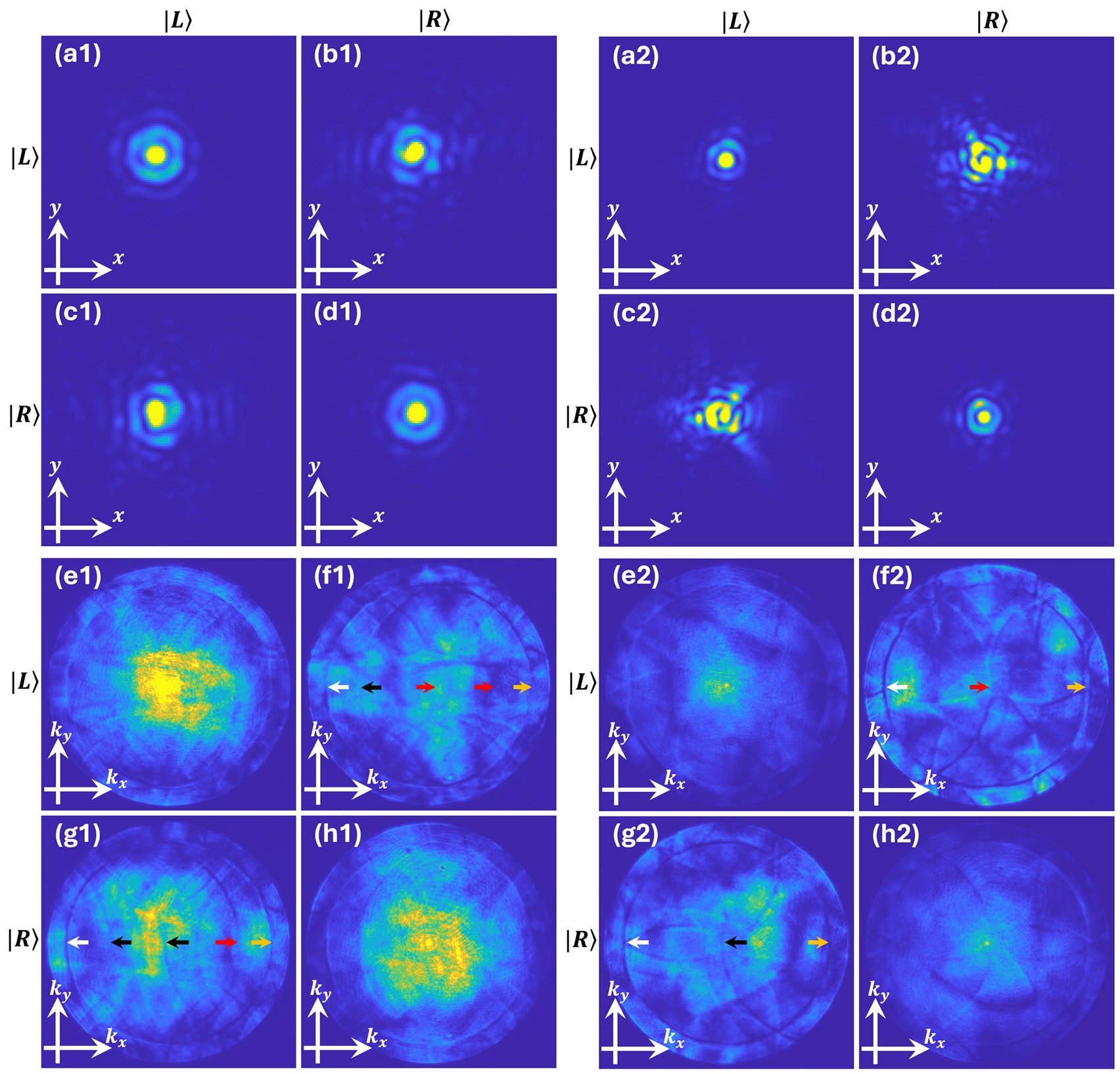}
    \caption{Results for Nonreciprocal Metasurfaces: (a1) - (h1) False chiral metasurface in a square lattice. (a2) - (h2) False chiral metasurface in a hexagonal lattice.
    Upper half represents real space results while the lower half shows $k$-space results with dispersion modes.
    The incident and measured polarizations are denoted on the left side and upper side of the figure, respectively, as $|L\rangle$ for LCP and $|R\rangle$ for RCP.
    The black and red arrows points to dispersion modes that produce polarization dependent SP propagation in the $-k_x$ and $+k_x$- directions, respectively.
    \label{fig: NR_ShP_Hex_RH}}
\end{figure}

Further, false chiral structures are being studied using our setup. For this purpose, we fabricate a metasurface comprising rectangular apertures ($150$ nm X $450$ nm) spatially rotated along the $x$ axis (shown in Fig. \ref{fig: Samples}b). As has been presented elsewhere\cite{fox2022generalized,fox2023topologically,ettapuram2025direct, gorodetski2009obs}, the rotation of the unit cell modifies the momentum conservation laws to yield 

\begin{equation}
k_{SP} = k_g + \sigma \Omega,
\label{momentum_eq}     
\end{equation}
where $\Omega = \frac{2\pi}{N\Lambda}$ is the spatial rotation frequency, $N$ is the number of periods within one aperture revolution and $\sigma = +1$ $\rightarrow$ RCP; $\sigma = -1$ $\rightarrow$ LCP. We chose to study two structure periods -- shorter period $\Lambda_{s} = 597\:nm$ and longer period $\Lambda_l = 995\:nm$ corresponding to the cases of $\sigma=\pm1$ in Eq. \ref{momentum_eq}. This choice ensures spin-dependent uni-directional coupling of incident light to plasmonic waves.
The measured results for the period $\Lambda_s$ with right handed rotation are presented in Fig. \ref{fig: NR_ShP_Hex_RH}a1-h1. Both the real and the $k$-space distributions are shown for four circular input/output states. We also investigated the metasurface of the period $\Lambda_s$ with an opposite sense of rotation and of the period $\Lambda_l$ with both left and right handed rotations; the results are presented in the Supplementary Material. 
A brief observation of the intensity distribution in the real space for the crossed polarizer/analyzer cases reveals a different result with comparison to the reciprocal cases. The spiral fringe suggests the appearance of a nonreciprocal phase that reduces the plasmonic vortex order to $l = \pm1$. The handedness change for the opposite incident polarizations cannot be explained by a constant phase added due to the structure and, therefore, is attributed to the time-reversal symmetry breaking in the medium. 

The origin of the nonreciprocity in this 1D spatially rotated structure can be explained by the excitation of an additional plasmonic mode of a particular circular polarization along one propagation direction.
\begin{figure}[htbp]
    \centering
    \includegraphics[width=0.45\textwidth]{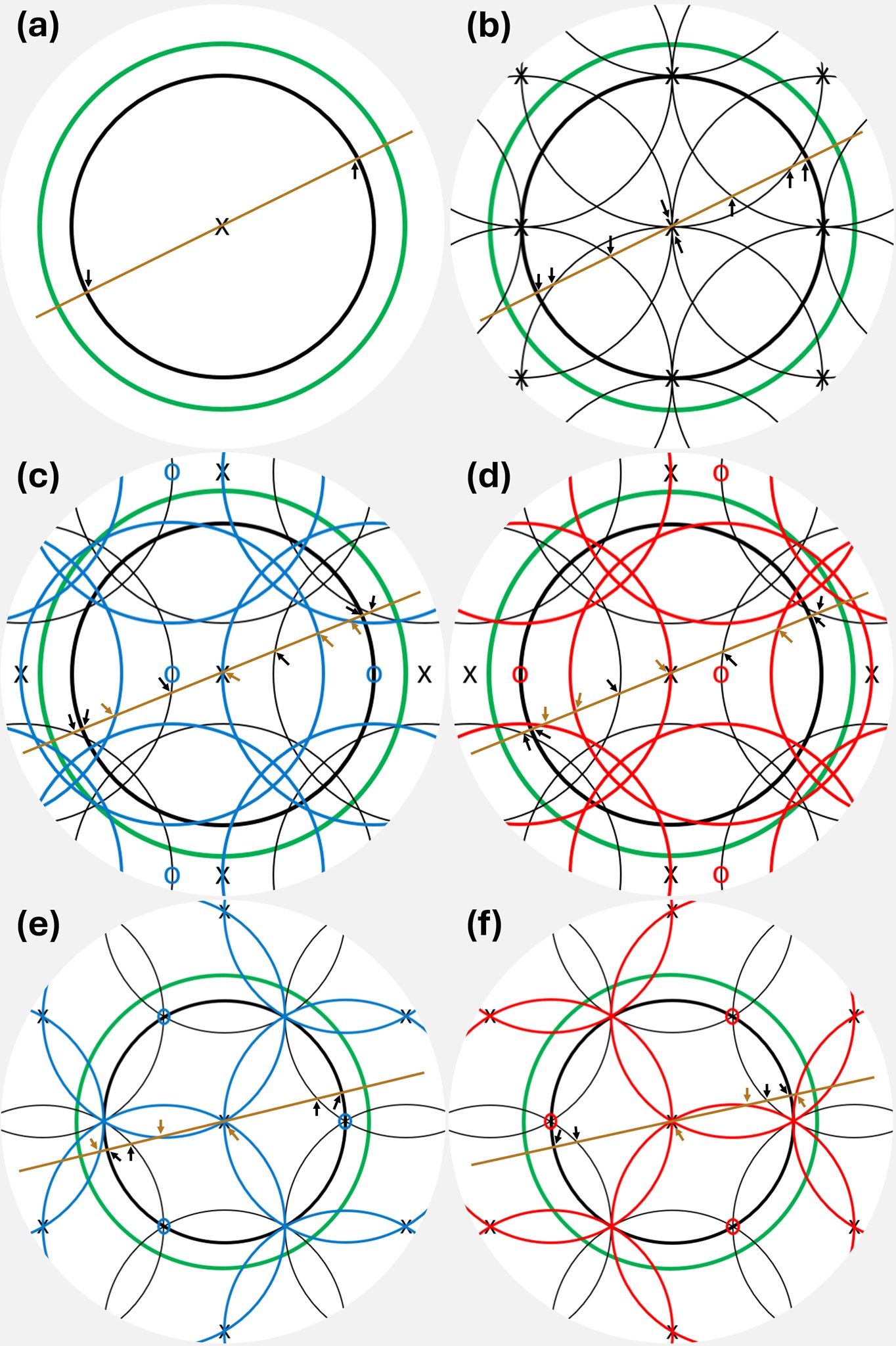}
    \caption{Schematic representation of SP dispersion modes in the k-space for: (a) flat surface, (b) HA grating, (c) Right Handed (RH) and (d) Left Handed (LH) Nonreciprocal Metasurface of Short Period ($\Lambda_{s}$) as in \ref{fig: Samples} b, 
    (e) Right Handed (RH) and (f) Left Handed (LH) Nonreciprocal Metasurface as in \ref{fig: Samples} c. 
    The black circles or arcs represent primary SP resonant dispersion mode and their Bloch replica. The red and blue circles or arcs represent polarization dependent SP dispersion modes.
    SP propagation along the brown diagonal line is considered. The black arrows point to the symmetrically distributed dispersion modes while the brown arrows point to the asymmetrically distributed dispersion modes in k-space.
    \label{fig: k_space_scheme}}
\end{figure}
The measured asymmetric distribution of the dispersion modes in the $k$-space is indicated by black and red arrows in Fig. \ref{fig: NR_ShP_Hex_RH}. The orange and white arrows point to the primary reciprocal (spatially symmetric) plasmonic modes excited in the systems. The origin of asymmetric modes lies in the excitation of the geometric phase due to a plasmonic spin-Hall effect in the space-variant grating \cite{bliokh2008coriolis,gorodetski2009observation}. Its worth noting that the high-NA illumination of the sample results in  polarization dependent SP modes across a wide angle range; therefore, they appear as arcs in the $k$-space (see the schemes in Fig. \ref{fig: k_space_scheme}). Accordingly, nonreciprocity of the structure can be assessed by counting the number of cross points on the radial line in a given direction with the SP modes. In Fig. \ref{fig: k_space_scheme}a,b, we demonstrate the iso-frequency surface (IFS) of the dispersion relation theoretically calculated for the flat surface and for the HA structure, respectively. The thick green circle represents the NA of the imaging system. The thick black circle represents the primary SP resonance line and is replicated (represented by thin black circles) due to the periodicity of the structure in two dimensions; however, the $k$-space stays symmetric. The brown line exemplifies an arbitrary radial direction. We note that it has an equal number of crossings with the plasmonic modes on both sides from the center. This symmetry holds for any direction and points to a system reciprocity. In Fig. \ref{fig: k_space_scheme}c,d, an IFS is shown for the previously-discussed metasurface comprising of a 1D aperture rotation for output $\left | R \right >$  and $\left | L \right >$ polarizations, respectively. Clearly, in this case the sample line drawn in the scheme reveals asymmetric crossings (indicated by brown arrows) on opposite sides, manifesting the nonreciprocity. As discussed previously\cite{ettapuram2025direct}, this structure is characterized by a time-reversal symmetry breaking in the $x$ direction resulting from the false chirality. Interestingly, when the radial line is rotated exactly to the $y$ direction, the number of mode-crossings on the opposite sides becomes equal. Accordingly, when exploring the IFS of a structure, one can predict the directions in which propagation of the SP waves is nonreciprocal.

An additional nonreciprocal structure of interest is a hexagonal aperture array, in which the rotation of the nanoslits occurs along three directions $0^{\circ}$, $60^{\circ}$ and $120^{\circ}$. The real-space distributions for the crossed polarization cases (Fig. \ref{fig: NR_ShP_Hex_RH}a2-d2) show an expected three-fold plasmonic beaming due to the coupling of geometric-phase orders to SPs\cite{fox2022generalized}. Moreover, a first order vortex appears in the center of the image. This suggests a structure nonreciprocity.
In the measured $k$-space images, the plasmonic modes appear in a form of a trillium flower that rotates with the polarization handedness (Fig. \ref{fig: NR_ShP_Hex_RH}e2-h2). The black and red arrows point to a couple of these polarization dependent modes. In the theoretical IFS diagram in Fig. \ref{fig: k_space_scheme}e,f, one can notice that the centers of the geometric phase orders are exactly located on the primary SP resonance. The sample line shown in the Figure manifests nonreciprocity. Nevertheless, it is possible to find three degenerate directions in which the system appears to be reciprocal for SP waves. These directions are defined by lines that are tangential to one of the circles exactly in the center of the $k$-space. 

\begin{figure}[htbp]
    \centering
    \includegraphics[width=0.45\textwidth]{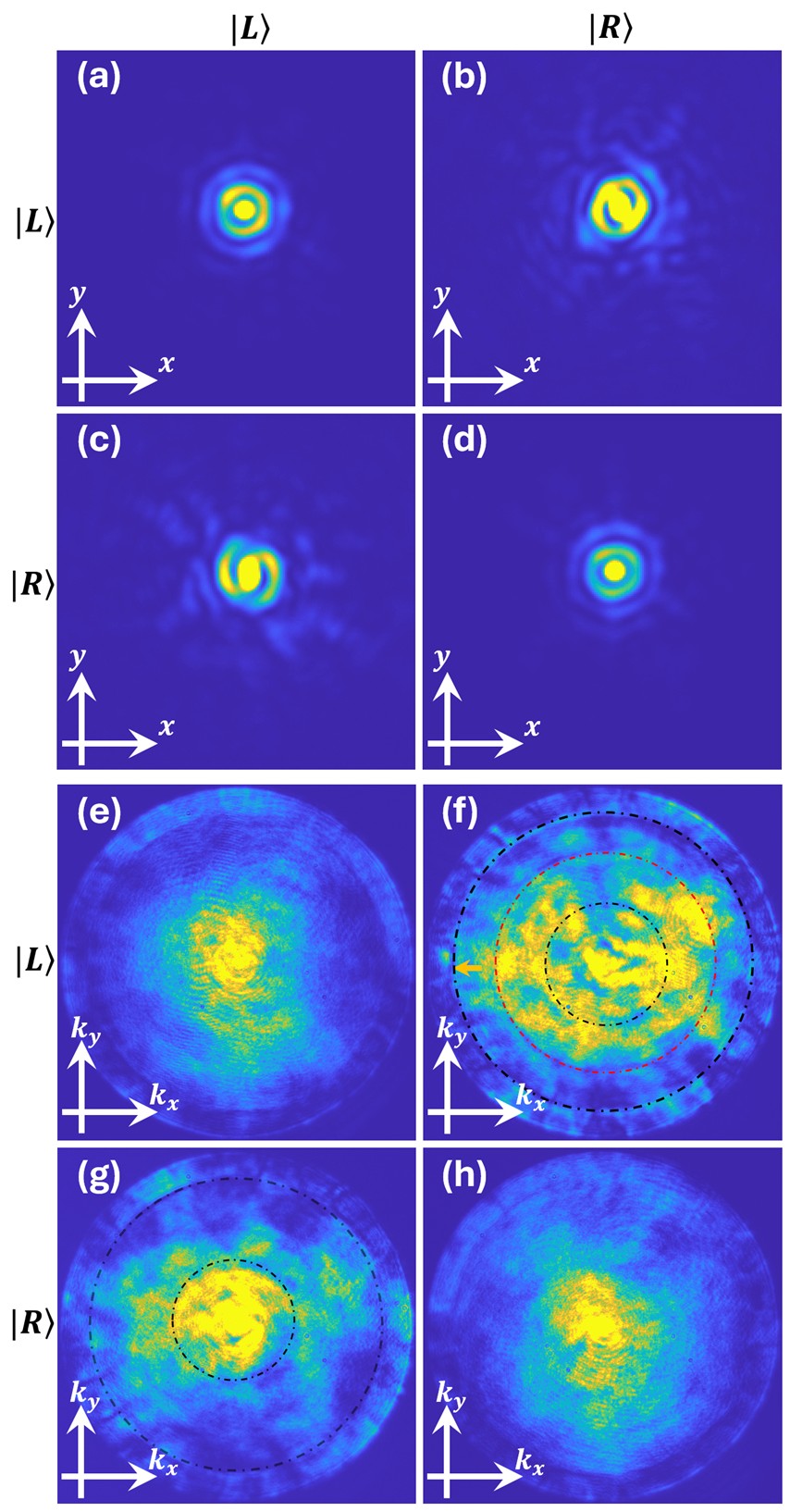}
    \caption{Results for true-chiral metasurface.
    (a) - (d) The real space results. 
    (e1) - (h1) The $k$-space results. 
    The incident and measured polarizations are denoted on the left side and upper side of the figure, respectively, as $|L\rangle$ for LCP and $|R\rangle$ for RCP.
    \label{fig: NR_Conical}}
\end{figure}

To this end, we conclude that our ILRM system detects structure nonreciprocity due to a false chirality and even visualizes its direction in the $k$-space. With this in mind, we wish to test a structure with a true chirality. We fabricate a radially periodic array of rotated rectangular apertures as shown in Fig. \ref{fig: Samples}d. The results are presented in Fig. \ref{fig: NR_Conical}. The apertures of the same angular orientation are placed in equally-distributed rings with a radial period of $\Lambda = 576\:nm$ and an azimuthal period (arc length) of $\Lambda = 603\:nm$. The relative rotation of the apertures in the consequent rings is $\pi/4$ (for N = 4). This structure is not superimposable with its own mirror image, so it is true-chiral by definition. The period of this structure was chosen similarly to the 1D grating, therefore Eq. \ref{momentum_eq} should hold here in radial coordinates and $k_{SP}>0$. Basically, the geometric phase in this case is built up as a positive or negative axicon. The $k$-space of this axicon can be constructed by revolving a one-dimensional dispersion ($\omega(k_r)$) about a vertical axis (and is described in detail in supplementary materials). In the IFS corresponding to our laser frequency, this sample is expected to have a diffraction order coinciding with $k_{SP}$ for only one input handedness whose dispersion mode will be at the center, while the opposite handedness produces a mode at a smaller radius compared to $k_{SP}$. So, in this structure, the additional momentum $\Omega$ can couple light to SPs in only one radial direction, outward or inward (depending on the light polarization and the period, $\Lambda$). The meta-axicon then can be designed to generate a converging or diverging SPs wave (compare the intensity distribution in Fig. \ref{fig: NR_Conical}f,g). Hence, this system can then be described as being ``radially'' nonreciprocal, since two waves propagating in opposite radial directions experience different adventures. Nevertheless, using our straight-forward approach, one can conclude that the system remains reciprocal along any chosen direction in the $k$-space. The real space images (see Fig. \ref{fig: NR_Conical}a1-d1) also points to its reciprocal nature as this sample produces second order vortex for crossed polarization cases similar to base interferograms.


\section{Conclusions}
In summary, we have experimentally demonstrated the detection and measurement of nonreciprocity and the determination of the directions of nonreciprocity, induced by a PB plasmonic metasurface featuring false chirality, using an ILRM setup.
The high-NA objective excitation produced SPs propagating in all directions in the 2D plane of the plasmonic sample, facilitating the opportunity to detect and study nonreciprocity in the entire 2D plane.
Nonchiral, false-chiral, and true-chiral samples were studied using the setup.
Reciprocal samples (nonchiral and true-chiral) produced a second order vortex in the real space under the crossed-polarization conditons and the presence of nonreciprocity in false-chiral samples was easily detected as a change in the topological charge of the produced vortex.
The nonreciprocity in false chiral samples is directly related to the asymmetric distribution of PB phase related disperion modes in the $k$-space, which was analyzed to determine the directions of nonreciprocity.
There are some discrete directions where the mode crossings are symmetric; except for these directions, all other directions are nonreciprocal in nature for false-chiral samples.
The reciprocal samples have a symmetric distribution of dispersion modes in any direction in the $k$-space.
This ILRM setup is a powerful tool for studying nonreciprocity and for mapping reciprocal or nonreciprocal modes present in a structure in the entire 2D plane.

\section*{Supplementary material}
The supplementary material consists of: theoretical representation of $k$-space image for axicon phase metasurface, additional results related to false chiral metasurfaces and the measurement of the quasi $k$-space diffraction orders.
\section*{AUTHOR DECLARATIONS}
\subsection*{Conflict of Interest}
The authors have no conflicts to disclose.
\subsection*{Author Contributions}
All authors have equally contributed to the work.
\begin{acknowledgments}
Authors acknowledge the Israeli Ministry of Innovation, Science and Technology for funding the research
We aknowledge fruitful discussions with prof. Shmuel Sternklar, Ariel University. 
\end{acknowledgments}
\section*{Data Availability Statement}
The data that support the findings of this study are available from the corresponding author upon reasonable request.

\bibliography{Ref}
\newpage
\begin{center}
    {\LARGE\bfseries Supplementary Material}\\[2em]
    {\Large\bfseries
    Inverted Leakage Radiation Microscopy for Measurement of Nonreciprocity Induced by a Plasmonic Metasurface with False Chirality
    }
\end{center}
\renewcommand{\thefigure}{S\arabic{figure}} 
\renewcommand{\thesubsection}{\Alph{subsection}} 
\setcounter{section}{0} 
\renewcommand{\thesection}{S\arabic{section}}

\section{Theoretical $k$-space image for Axicon phase metasurface}
\begin{figure}[htbp]
    \centering
    \includegraphics[width=0.45\textwidth]{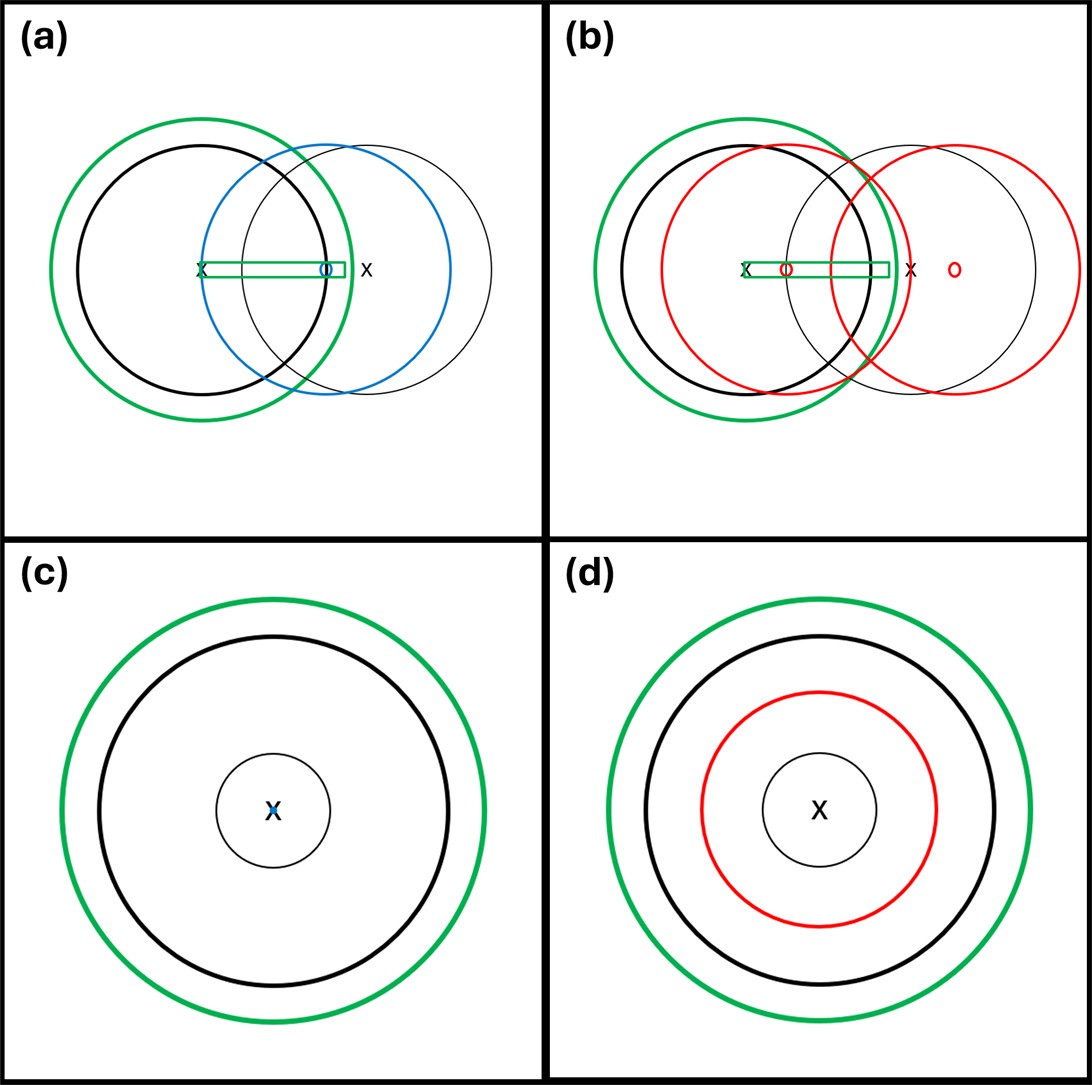}
    \caption{Theoretical representation of $k$-space SP dispersion modes in the $+k_x$ direction for a metasurface of rectangular apertures rotated along $x$-direction with a period of $576 nm$. An aperture revolution completes in 4 periods (N=4). (a) \& (c) LCP related modes, (b) \& (d) RCP related modes.
    The black circles or arcs represent primary SP resonant dispersion mode and their Bloch replica. The red and blue circles or arcs represent polarization dependent SP dispersion modes.
    \label{fig: k_space_scheme_Axicon}}
\end{figure}
The 1D dispersion scheme that corresponds to diffraction orders in the $+k_x$ - direction is shown in Fig. \ref{fig: k_space_scheme_Axicon}a \& b for the $|L\rangle$ and $|R\rangle$ polarizations, respectively.
The green box shows dispersion modes that are strictly in the $+k_x$ direction and within the numerical aperture of the objective lens.
Fig. \ref{fig: k_space_scheme_Axicon}c \& d show the $k$-space scheme for the axicon once the modes in the green box revolves about a vertical axis at the center.
The RCP dispersion mode in Fig. \ref{fig: k_space_scheme_Axicon}b is at $+\frac{2}{3} k_{sp}$ shown in red color in the green box propagates in the $+x$ direction and hence will have an outward propagation. i.e, the red dispersion mode in Fig. \ref{fig: k_space_scheme_Axicon}c will propagate outward from the center and spread.
From Fig. \ref{fig: k_space_scheme_Axicon}a, the polarization dependent dispersion mode is at the center, i.e., at normal incidence angle. 
Similarly, the LCP dispersion mode in Fig. \ref{fig: k_space_scheme_Axicon}d will propagate inward and hence focuses towards the center. This spreading and focusing for different polarizations is clearly visible when the intensity distribution in the $k$-space is observed (see Fig. 6f \& g in the main text).

\section{False Chiral Experiments}
\begin{figure}[htbp]
    \centering
    \includegraphics[width=0.4\textwidth]{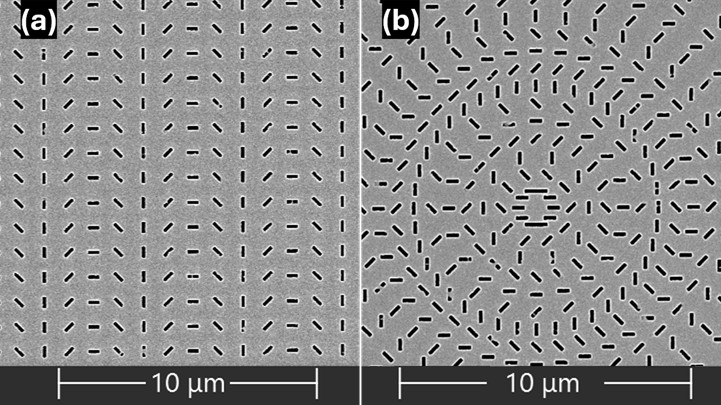}
    \caption{SEM image of false-chiral metasurface
    (a) of longer period ($\Lambda_{l}$) with spatially rotated rectangular apertures along the $x$ axis in a square lattice.
    (b) in a circular lattice; a  single line of 1D false-chiral metasurface is rotated in 180 degrees.
    \label{fig: Samples1}}
\end{figure}
\begin{figure}[htbp]
    \centering
    \includegraphics[width=0.9\textwidth]{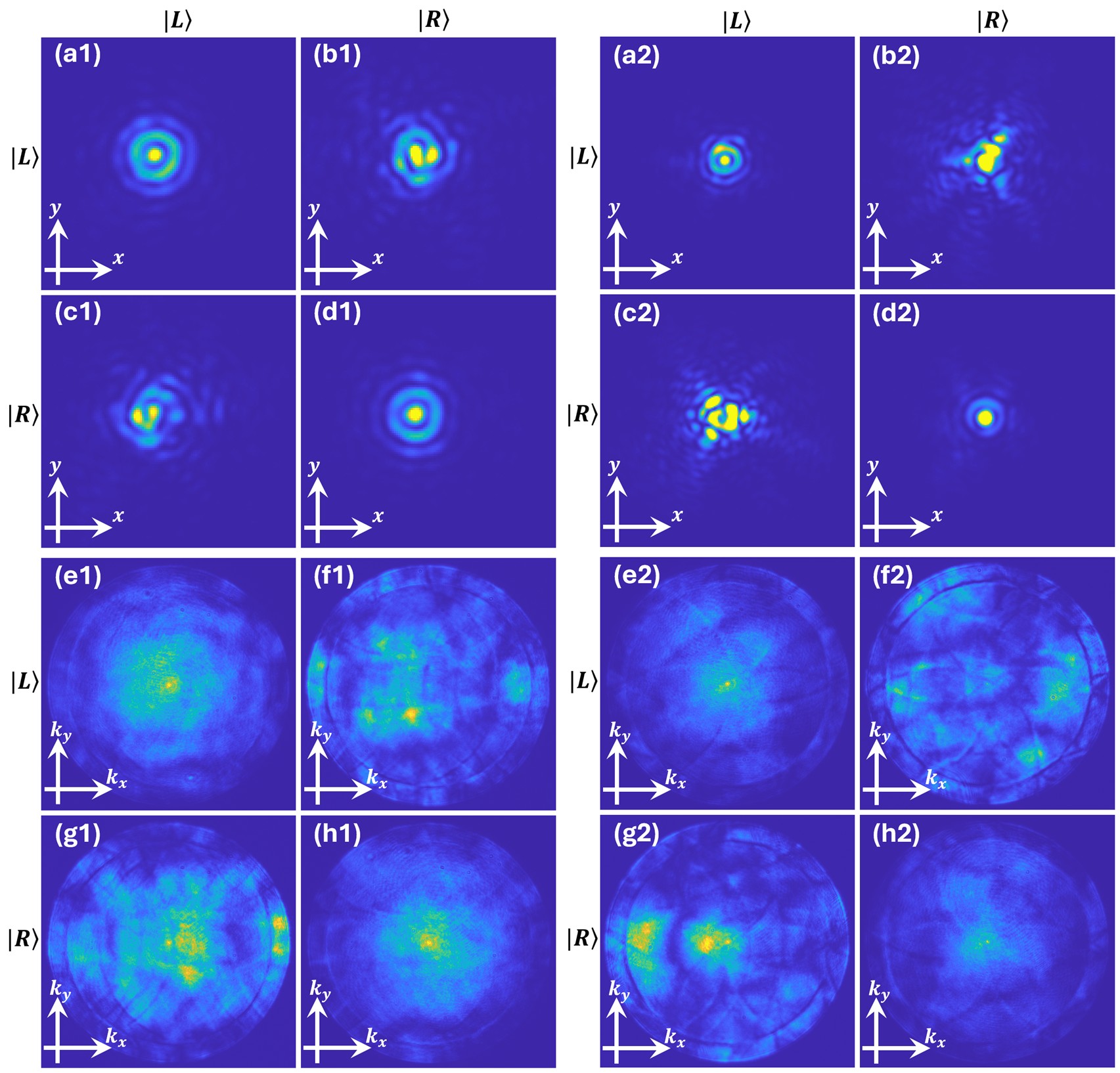}
    \caption{Results for Left Handed nonreciprocal metasurfaces: (a1) - (h1) False chiral metasurface in a square lattice. (a2) - (h2) False chiral metasurface in a hexagonal lattice.
    Upper half represents real space results while the lower half shows the $k$-space results.
    The incident and measured polarizations are denoted on the left side and upper side of the figure, respectively, as $|L\rangle$ for LCP and $|R\rangle$ for RCP.
    \label{fig: NR_ShP_Hex_LH}}
\end{figure}
\begin{figure}[htbp]
    \centering
    \includegraphics[width=0.9\textwidth]{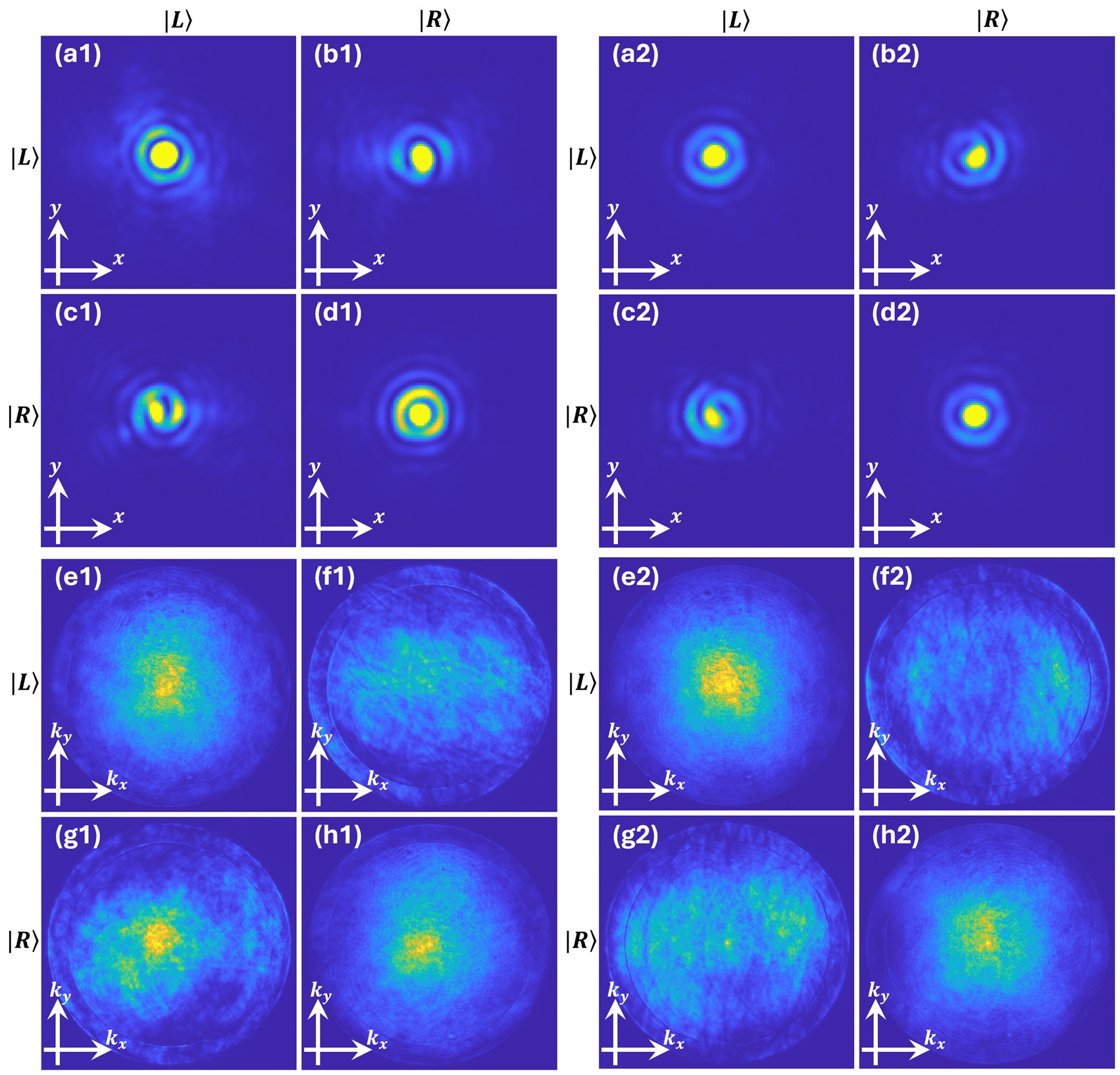}
    \caption{Results for (a1) - (h1) Right handed and (a2) - (h2) Left Handed Nonreciprocal Metasurfaces:  False-chiral metasurface in a square lattice with longer period.
    Upper half represents real space results while the lower half shows $k$-space results.
    The incident and measured polarizations are denoted on the left side and upper side of the figure, respectively, as $|L\rangle$ for LCP and $|R\rangle$ for RCP.
    \label{fig: NR_LoP_RH_LH}}
\end{figure}
\begin{figure}[htbp]
    \centering
    \includegraphics[width=0.9\textwidth]{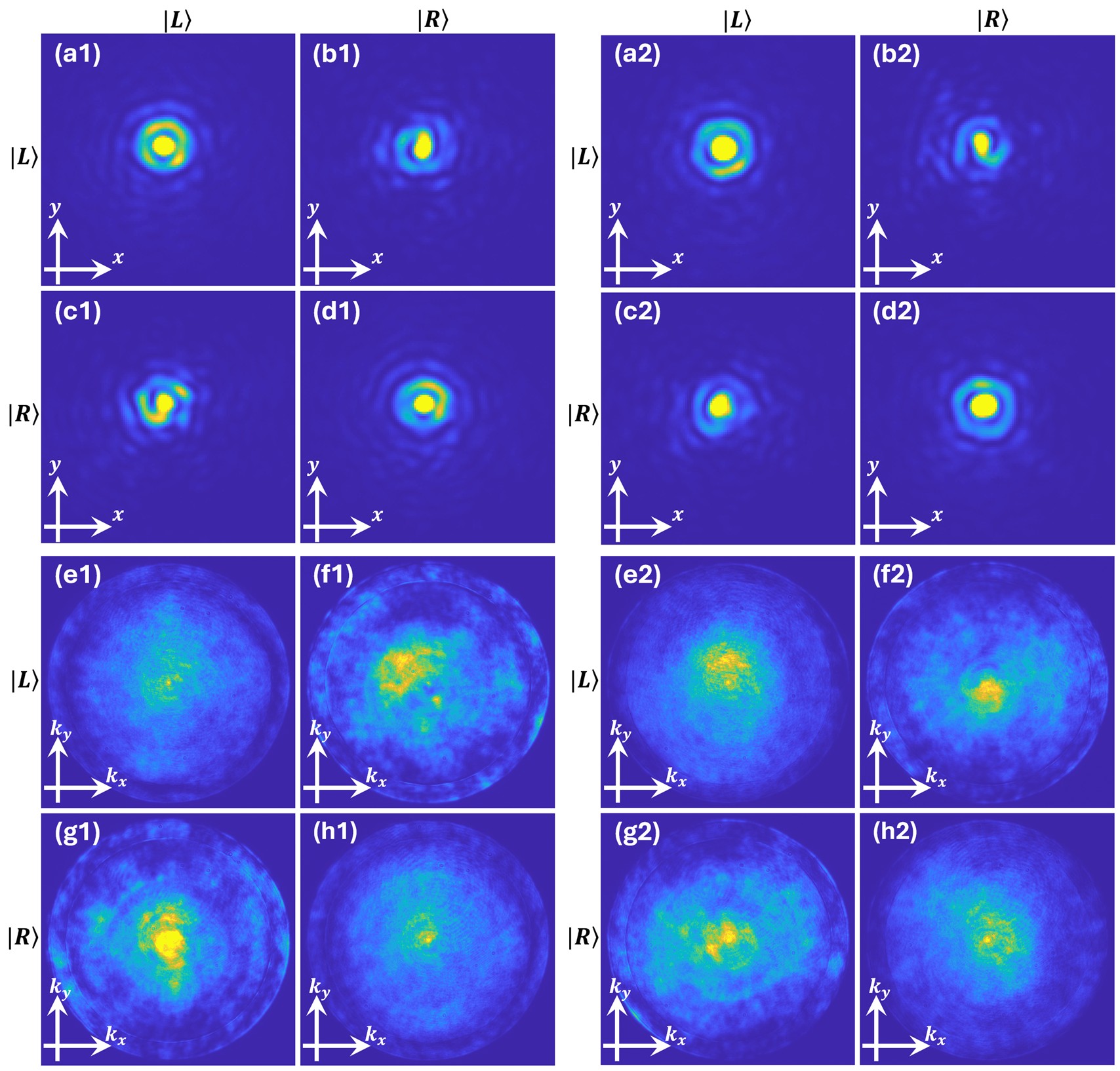}
    \caption{Results for false-chiral metasurfaces of circular lattice as shown in \ref{fig: Samples1} (b).
    Upper half represents real space results while the lower half shows $k$-space results.
    The incident and measured polarizations are denoted on the left side and upper side of the figure, respectively, as $|L\rangle$ for LCP and $|R\rangle$ for RCP.
    \label{fig: NR_2D_Circ}}
\end{figure}

Fig. \ref{fig: NR_ShP_Hex_LH} shows the results for left handed metasurfaces corresponding to the right handed nonreciprocal metasurfaces studied in Fig. 4 in the main text.
Fig. \ref{fig: NR_LoP_RH_LH} shows the results for false-chiral metasurfaces in a square lattice (see Fig. \ref{fig: Samples1}a) with a Longer Period, $\Lambda_l = 995\:nm$, corresponding to the case of $\sigma=+1$ in Eq. 1 in the main text.
Fig. \ref{fig: NR_2D_Circ} shows the results for false-chiral metasurfaces in circular lattices, as shown in Fig. \ref{fig: Samples1}b.
Real space measurements in crossed-polarization conditions show that the characteristic double vortex has changed to an order of $\pm 1$, suggesting the presence of nonreciprocity.
The dispersion modes in the $k$-space also show an asymmetric distribution related to the induced nonreciprocity.

\section{Quasi-$k$-space diffraction}
\begin{figure}[htbp]
    \centering
    \includegraphics[width=0.45\textwidth]{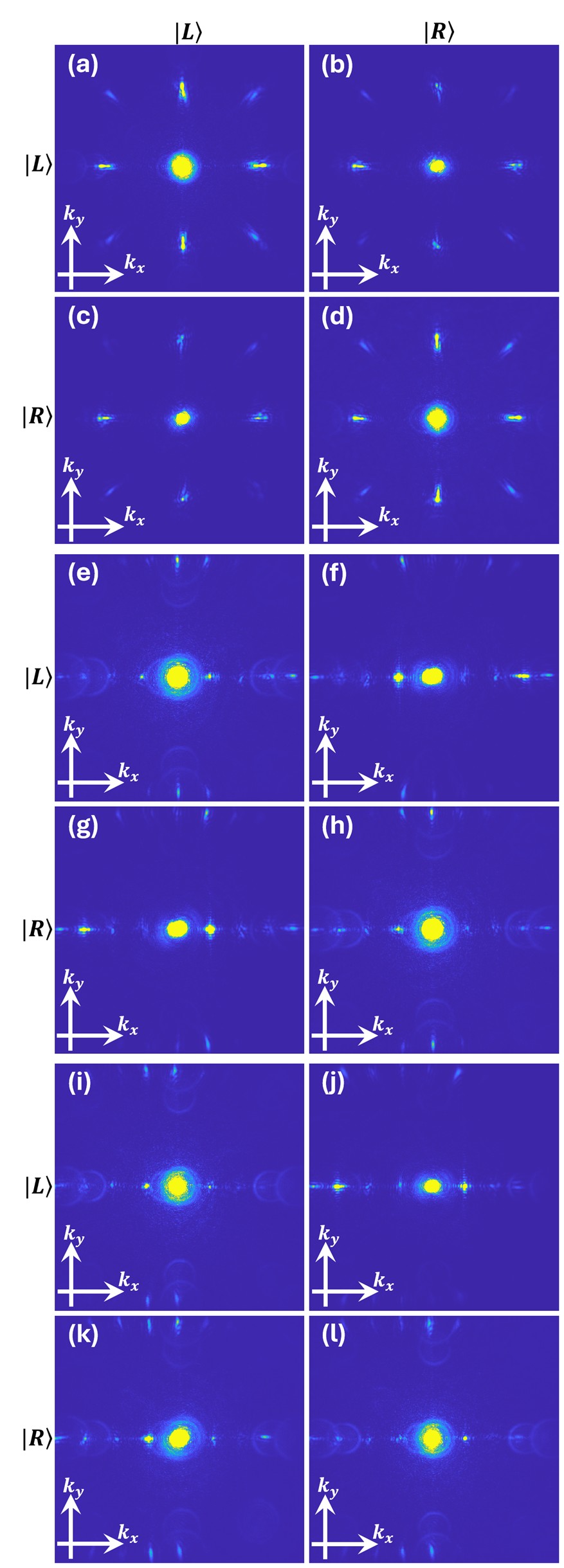}
    \caption{Diffraction orders for (a) - (d) Hole Array,  (e) - (h) right handed and (i) - (l) left handed 1D shorter period nonreciprocal metasurfaces in a square lattice.
    The incident and measured polarizations are denoted on the left side and upper side of the figure, respectively, as $|L\rangle$ for LCP and $|R\rangle$ for RCP.
    \label{fig: Diffraction_HA_ShP}}
\end{figure}

\begin{figure}[htbp]
    \centering
    \includegraphics[width=0.45\textwidth]{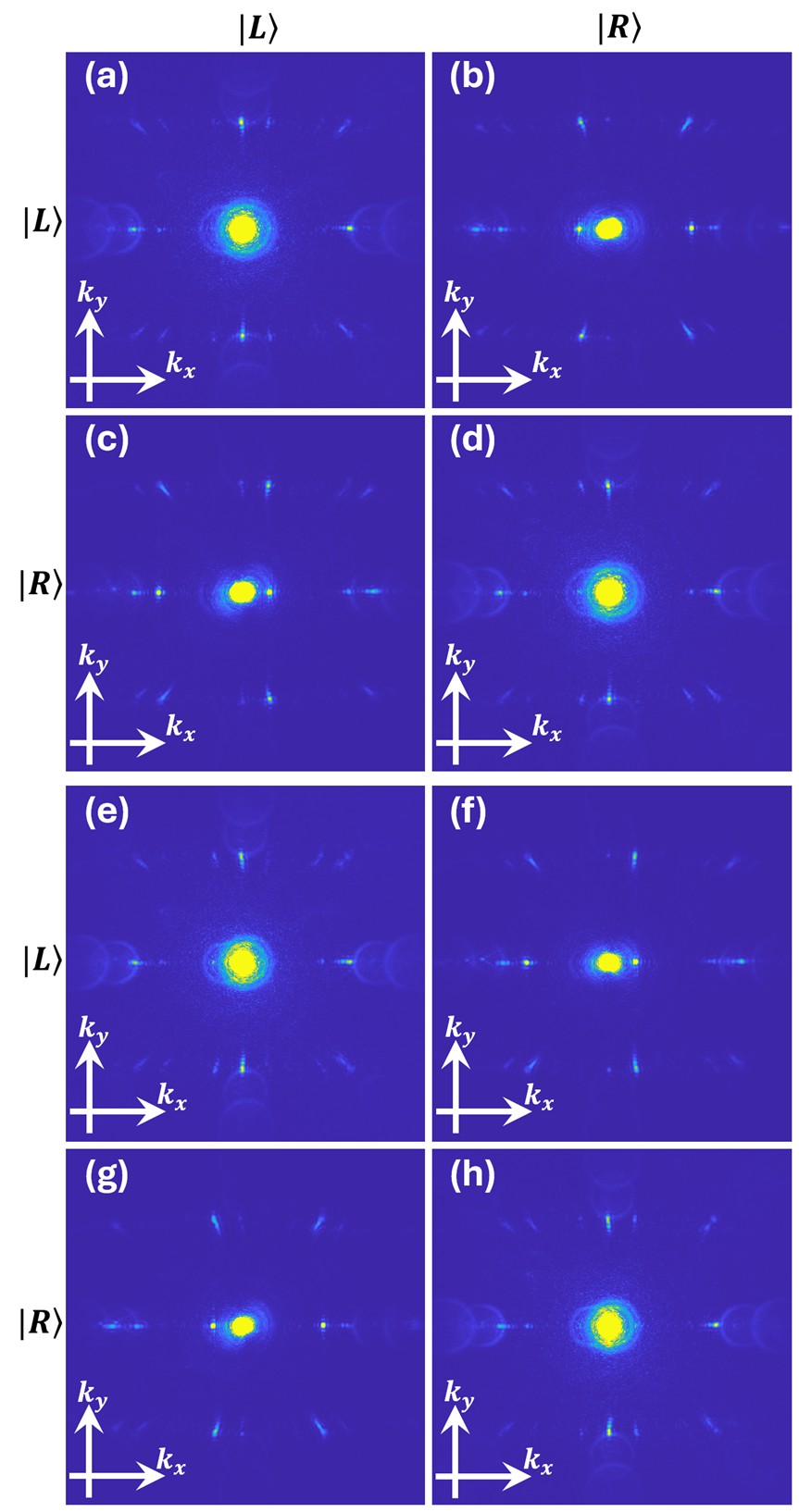}
    \caption{Diffraction orders for (a) - (d) right handed and (e) - (h) the left handed 1D longer period nonreciprocal metasurfaces in a square lattice.
    The incident and measured polarizations are denoted on the left side and upper side of the figure, respectively, as $|L\rangle$ for LCP and $|R\rangle$ for RCP.
    \label{fig: Diffraction_LoP}}
\end{figure}

\begin{figure}[htbp]
    \centering
    \includegraphics[width=0.45\textwidth]{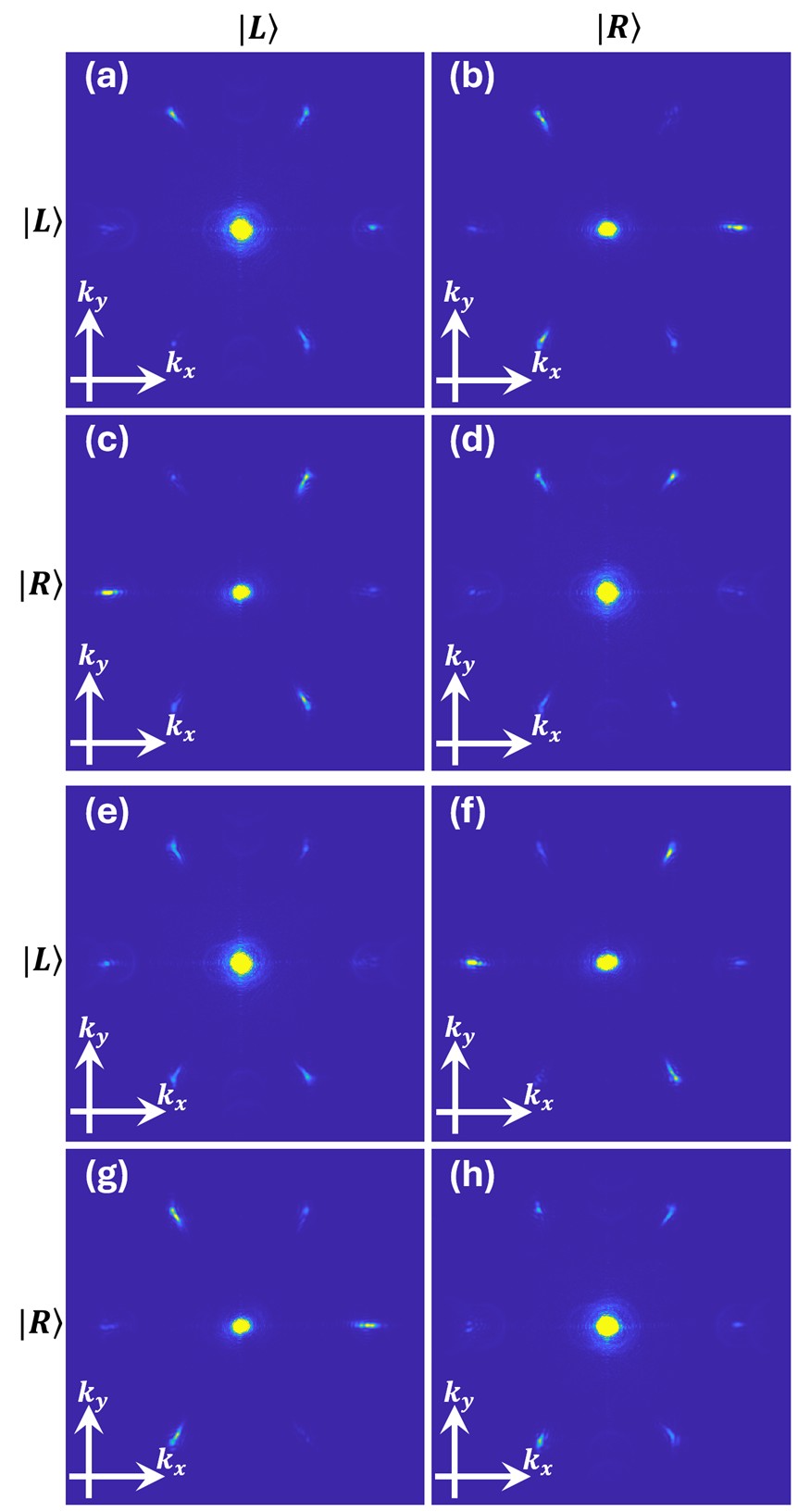}
    \caption{Diffraction orders for (a) - (d) right handed and (e) - (h) the left handed nonreciprocal metasurfaces in a hexagonal lattice.
    The incident and measured polarizations are denoted on the left side and upper side of the figure, respectively, as $|L\rangle$ for LCP and $|R\rangle$ for RCP.
    \label{fig: Diffraction_Hex}}
\end{figure}

\begin{figure}[htbp]
    \centering
    \includegraphics[width=0.45\textwidth]{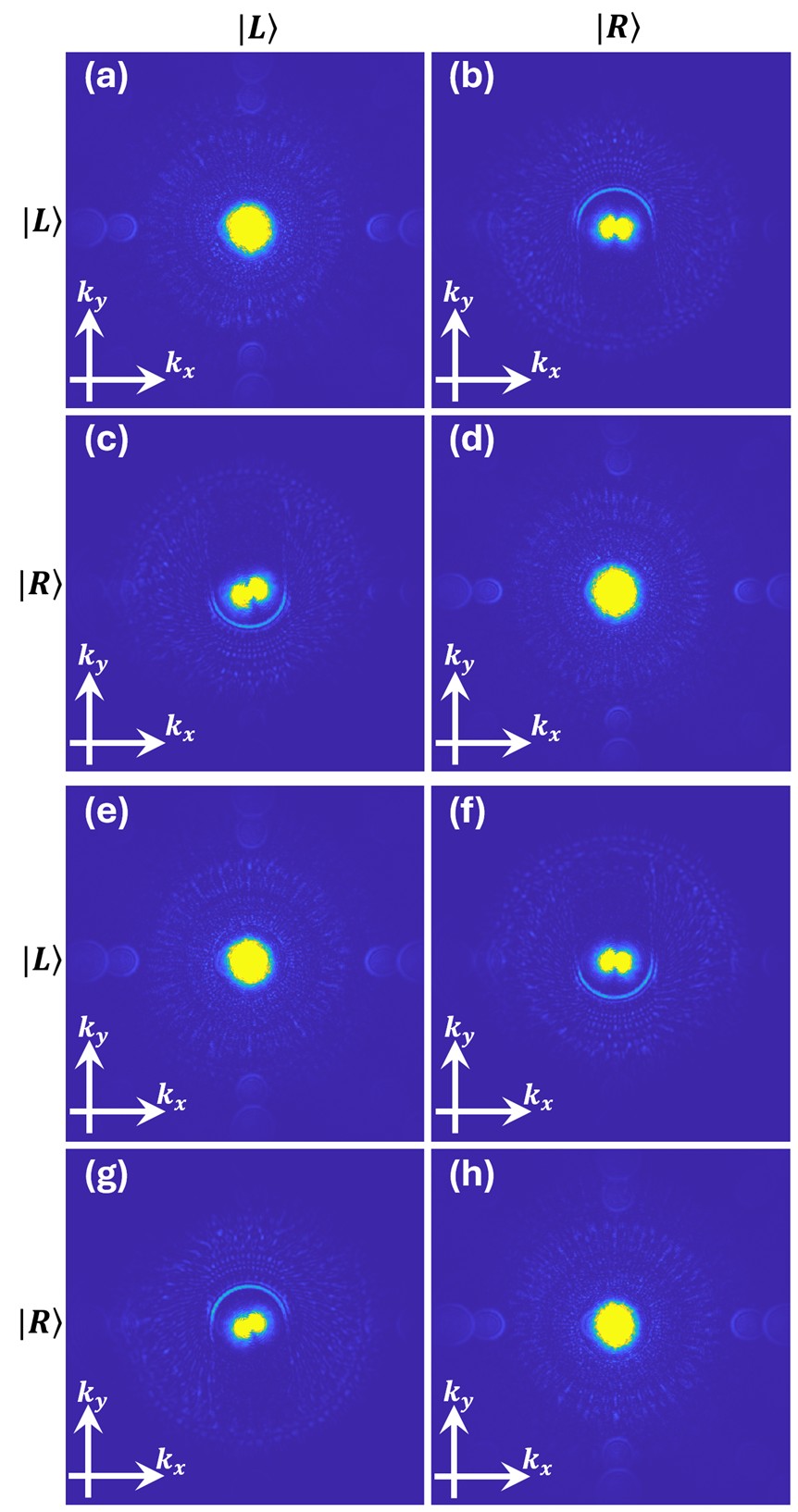}
    \caption{Diffraction orders for (a) - (d) right handed and (e) - (h) the left handed nonreciprocal metasurfaces of circular lattice.
    The incident and measured polarizations are denoted on the left side and upper side of the figure, respectively, as $|L\rangle$ for LCP and $|R\rangle$ for RCP.
    \label{fig: Diffraction_circ}}
\end{figure}

The diffraction orders for the given structures were measured in a quasi-$k$-space measurement in the inverted leakage radiation microscropy setup. For this purpose, the sample was placed in a slightly pre-focal position of the objective lens, Obj2. Fig. \ref{fig: Diffraction_HA_ShP} shows the diffraction orders for a Hole Array (see Fig. 2a in main text), and 1D shorter period false chiral sample (see Fig. 2b in main text).
Fig. \ref{fig: Diffraction_LoP} shows the diffraction orders for a 1D longer period false chiral sample in square lattice (see Fig. \ref{fig: Samples1}a).
Fig. \ref{fig: Diffraction_Hex} shows the diffraction orders for a false chiral sample in a hexagonal lattice (see Fig. 2c in main text).
Fig. \ref{fig: Diffraction_circ} shows the diffraction orders for a false chiral sample in circular lattice (see Fig. \ref{fig: Samples1}b).

\end{document}